\documentclass{PoS}
\usepackage{epsfig}
\usepackage{amsmath}
\usepackage{amssymb,amsfonts}
\usepackage{graphicx}
\usepackage{graphics}
\usepackage{amsthm}
\usepackage{graphicx}
\usepackage{multirow}

\newcommand{\cF}{{\cal F}}
\newcommand{\cFb}{{\overline{\cal F}}}
\newcommand{\cD}{{\cal D}}

\newcommand{\cQ}{{\cal Q}}
\newcommand{\cU}{{\cal U}}

\newcommand{\cUb}{{\overline{\cal U}}}
\newcommand{\Tr}{{\rm Tr\;}}

\newcommand{\hatbmu}{\widehat{\boldsymbol {\mu}}}

\newcommand{\vn}{ {\bf n} }

\def\({\left(}
\def\){\right)}
\def\[{\left[}
\def\]{\right]}

\def\nn{\nonumber}
\def\bec{\begin{center}}
\def\eec{\end{center}}
\def\beq{\begin{equation}}
\def\eeq{\end{equation}}
\def\bea{\begin{eqnarray}}
\def\eea{\end{eqnarray}}

\usepackage{subfigure}

\title{Eigenvalue spectrum of lattice $\mathcal{N}=4$ super Yang-Mills}

\ShortTitle{Eigenvalue spectrum of lattice $\mathcal{N}=4$ super Yang-Mills}

\author{\speaker{David J. Weir} \\
Department of Physics and Helsinki Institute of Physics,\\
PL 64 (Gustaf H\"allstr\"omin katu 2), FI-00014 University of
Helsinki, Finland \\
E-mail: \email{david.weir@helsinki.fi}
}

\author{Simon Catterall \\
Department of Physics, Syracuse University, Syracuse, NY 13244, USA \\
E-mail: \email{smc@physics.syr.edu}
}

\author{Dhagash Mehta \\
Department of Mathematics, North Carolina State University, Raleigh, NC 27695-8205, USA.
E-mail: \email{dbmehta@ncsu.edu}
}

\abstract{We present preliminary results for the eigenvalue spectrum of four-dimensional ${\cal N}=4$ super Yang-Mills theory on the lattice. In particular,
by studying the the spectral density a measurement of the anomalous dimension is made and found to be consistent with zero.
}

\FullConference{31st International Symposium on Lattice Field Theory - LATTICE 2013\\
		July 29 - August 3, 2013\\
		Mainz, Germany}

\begin{document}

\section{Introduction}\vspace{-0.2in}
In recent years a supersymmetric lattice regularization of ${\cal N}=4$ super Yang-Mills has been developed \cite{Catterall:2009it, Catterall:2011pd, Catterall:2013roa}. A preliminary
numerical exploration of the phase diagram was conducted in \cite{Catterall:2012yq} and evidence that the theory (at least for gauge group $U(2)$) was not subject to a
sign problem was reported in \cite{Catterall:2011aa,Mehta:2011ud,Galvez:2012sv}). In this article we analyze the spectrum of the fermion operator in this theory. The spectrum is important since it can
yield important information on both  the mass anomalous dimension of the theory and the fluctuations in the phase of the Pfaffian that results after one
integrates over the fermions in the theory. In this note we present preliminary results from an analysis at several values of the 't Hooft parameter, the
scalar mass (included to regularize the flat directions) and several lattice sizes.

The lattice theory results from discretization of a {\it twisted} form of the super Yang Mills theory. While in  flat space it  is equivalent to the usual theory the fields
appearing in the twisted model appear quite different; the twisted fermions appear as antisymmetric tensor components of a K\"{a}hler-Dirac field and the bosons fields
are packaged into 5 complexified gauge fields. Furthermore, the natural lattice associated with the discrete theory is the $A_4^*$ lattice whose basis vectors correspond
to the fundamental weights of $SU(5)$. All fields are associated to links in this lattice.
The action for this theory is
\bea
\label{eq:action-on-a4-star}
S &=& \frac{N}{2\lambda}\sum_{\vn, a,b,c,d,e} ~\Big\{ \cQ ~\Tr \Big[-i\chi_{ab} \cD^{(+)}_a \cU_b(\vn) - \eta(\vn) \Big(i\cD^{\dagger(-)}_a \cU_a(\vn) - \frac{1}{2}d(\vn) \Big)\Big]\nn \\
&& - \frac{N}{8\lambda} \Tr \epsilon_{abcde} \chi_{de}(\vn + \hatbmu_a + \hatbmu_b + \hatbmu_c) \cD^{\dagger(-)}_{c} \chi_{ab}(\vn + \hatbmu_c)\Big\}~.
\eea
where the lattice field strength is given by
\beq
\cF_{ab}(\vn) \equiv -\frac{i}{g}\cD^{(+)}_a \cU_b(\vn) = -\frac{i}{g}\Big(\cU_a(\vn) \cU_b(\vn + \hatbmu_a) - \cU_b(\vn) \cU_a(\vn + \hatbmu_b)\Big).
\eeq
and the covariant difference operators appearing in this expression
are given by
\bea
\cD_c^{(+)} f(\vn) &=& \cU_c(\vn) f(\vn + \hatbmu_c) - f(\vn) \cU_c(\vn), \\
\cD_c^{(+)} f_d(\vn) &=& \cU_c(\vn) f_d(\vn + \hatbmu_c) - f_d(\vn) \cU_c(\vn + \hatbmu_d), \\
\cD_c^{\dagger(-)} f_c(\vn) &=& f_c(\vn)\cU^{\dagger}_c(\vn) - \cU^{\dagger}_c(\vn - \hatbmu_c) f_c(\vn - \hatbmu_c), \\
\cD_c^{\dagger(-)} f_{ab} (\vn) &=& f_{ab}(\vn) \cU^{\dagger}_c(\vn - \hatbmu_c) - \cU^{\dagger}(\vn + \hatbmu_a + \hatbmu_b - \hatbmu_c) f_{ab}(\vn - \hatbmu_c).
\eea
The action of the scalar supercharge on the fields in the twisted theory is given by:
\begin{eqnarray}
\cQ \cU_{a} &=& \psi_{a} \\
\cQ \psi_a &=& 0 \\
\cQ \cUb_a &=& 0 \\
\cQ \chi_{ab} &=& \cFb_{ab} \\
\cQ \eta &=& d \\
\cQ d &=& 0
\end{eqnarray}
Supersymmetric invariance of the $\cQ$-exact part of the action then follows from the nilpotent property of $\cQ$ while an exact lattice Bianchi
identity ensures the $\cQ$ invariance of the $\cQ$-closed term. We simulate this theory by first integrating out the twisted fermions and using the RHMC algorithm to
reproduce the resultant (phase quenched) Pfaffian. We have implemented an Omelyan multistep integrator to improve the efficiency of the update and
employ a GPU accelerated multimass solver for speedup when available.

In practice we have introduced a small mass shift in the fermion operator to avoid an exact zero mode (we use periodic boundary
conditions in all directions) and have added an additional scalar mass
term to regulate the flat directions in the model of the form
\beq
\Delta S=\mu^2\sum_{x,a}\left(\frac{1}{N}\Tr\left(\cUb_a(x)\cU_a(x)\right)-1\right)^2\eeq
We have conducted simulations for a range of 't Hooft coupling $\lambda$, scalar mass $\mu$ and lattice size $L$.

\section{Results}\vspace{-0.2in}
We simulate with $\delta t=0.2$ (and a trajectory length of $1$) and carry out measurements every ten trajectories. For each volume and parameter choice, we measure the lowest 200
eigenvalues of $D^\dagger D$ using the ARPACK package~\cite{ARPACK}. This implements a Krylov subspace technique for numerical diagonalization called the implicitly restarted Arnoldi algorithm. The
eigenvalues of our operator come in real pairs, so we obtain 100 distinct eigenvalues. For $L=8$, each call to ARPACK takes about an hour to complete. The total computer time used to generate the configurations and analyse the eigenspectra presented here is approximately a hundred thousand hours. Results of simulations from ten independently thermalized lattices were used for each parameter choice.

\begin{figure}
\begin{center}\includegraphics[trim=1 1 1 1,clip=true,width=0.7\textwidth]{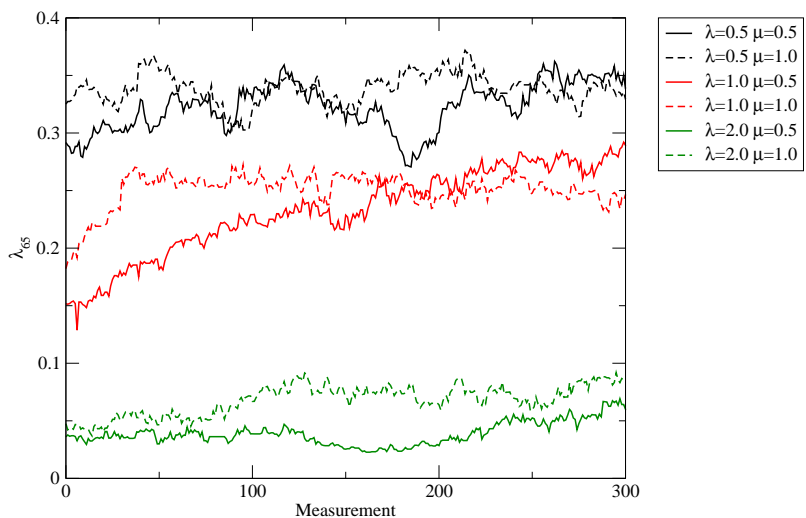}\end{center}
\caption{\label{fig:timeseries} Time series of eigenvalue
  measurements of $D^\dagger D$ for $L=6$ and various $\lambda$ and
  $\mu$. The eigenvalue measurements thermalize relatively quickly, although $\lambda=1.0$, $\mu=0.5$ is noticeably slower than the others.}
\end{figure}

In most cases, the eigenspectrum seems to thermalize relatively quickly, as shown in
Figure~\ref{fig:timeseries}. The exception seems to be $\lambda=1$, $\mu=0.5$, although other quantities do not exhibit the same issue; its precise origin is unclear.

To be conservative, we discard the first fifty measurements
from each time series and then bin based on an autocorrelation time
measurement for the rest. The number of (unbinned) measurements varied
from at least 5000
for $L=2$ to $L=6$ down to around 1000 for $L=8$. Ten
independently thermalized ensembles were used for each parameter
choice and volume to achieve the required statistics.

\begin{figure}
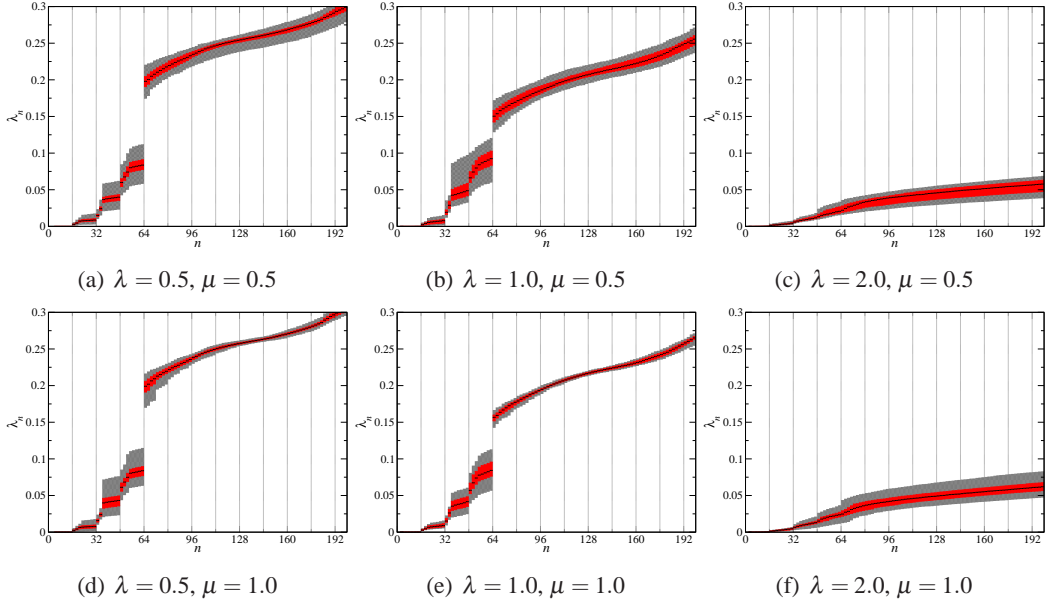

\begin{center}
\subfigure[$\lambda=0.5$, $\mu=0.5$]{\includegraphics[trim=1 1 1 1,clip=true,width=0.3\textwidth]{L8-lam0.5-mu0.5-quartiles.eps}}
\subfigure[$\lambda=1.0$, $\mu=0.5$]{\includegraphics[trim=1 1 1 1,clip=true,width=0.3\textwidth]{L8-lam1.0-mu0.5-quartiles.eps}}
\subfigure[$\lambda=2.0$, $\mu=0.5$]{\includegraphics[trim=1 1 1 1,clip=true,width=0.3\textwidth]{L8-lam2.0-mu0.5-quartiles.eps}}
\subfigure[$\lambda=0.5$, $\mu=1.0$]{\includegraphics[trim=1 1 1 1,clip=true,width=0.3\textwidth]{L8-lam0.5-mu1.0-quartiles.eps}}
\subfigure[$\lambda=1.0$, $\mu=1.0$]{\includegraphics[trim=1 1 1 1,clip=true,width=0.3\textwidth]{L8-lam1.0-mu1.0-quartiles.eps}}
\subfigure[$\lambda=2.0$, $\mu=1.0$]{\includegraphics[trim=1 1 1 1,clip=true,width=0.3\textwidth]{L8-lam2.0-mu1.0-quartiles.eps}}
\end{center}
\caption{\label{fig:quartiles} Full eigenvalue distribution for $L=8$
  and various $\lambda$ and $\mu$. The median eigenvalue is shown in black, and
  the red and grey bands identify the quartiles in the eigenvalue
  distribution. Our further analyses in this paper start with the 65th
eigenvalue shown in these plots.}
\end{figure}

\begin{figure}
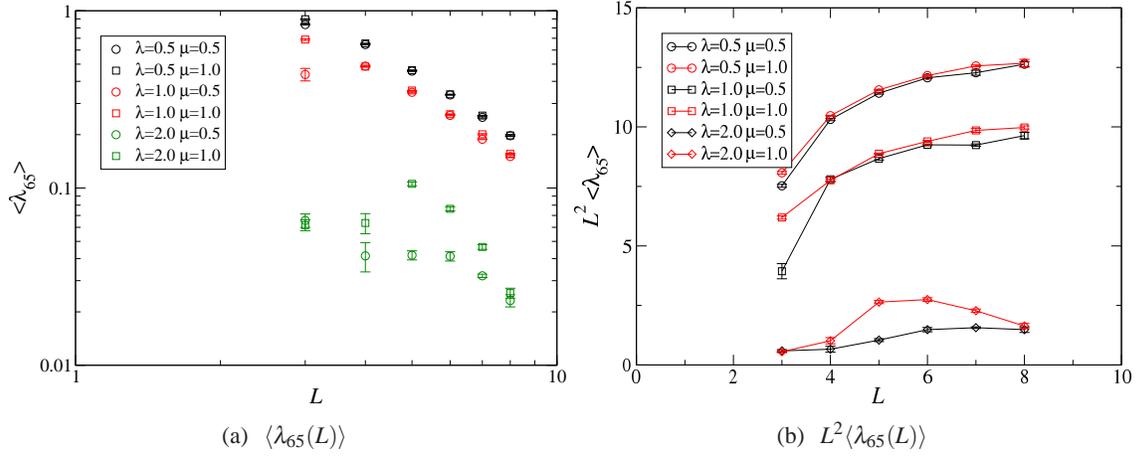

\begin{center}
\subfigure[\label{fig:loglogev1} $\langle
  \lambda_{65}(L) \rangle$ ]{\includegraphics[trim=1 1 1 1,clip=true,width=0.49\textwidth]{loglogev65.eps}}
\subfigure[\label{fig:lsqev65} $L^2\langle \lambda_{65}(L) \rangle$]{\includegraphics[trim=1 1 1 1,clip=true,width=0.49\textwidth]{lsq.eps}}
\end{center}
\caption{Plots of the scaling with volume for $\lambda_{65}$ at various $\lambda$ and $\mu$. In (a) the eigenvalues are shown, unscaled, on a doubly logarithmic plot. In the linear plot (b) the eigenvalues have been scaled by $L^2$.}
\end{figure}

Figure~\ref{fig:quartiles} shows typical eigenspectra for our
parameter choices at the largest volume, $L=8$. Amongst other finer
details, clear jumps can be seen at the 16th, 32nd, 48th and
64th eigenvalues. Furthermore, the scaling with volume of the first
64 eigenvalues is somewhat different from subsequent ones (see
Figure~\ref{fig:loglogev1} for the clear volume scaling seen from
$\lambda_{65}$).

These 64 modes can be analysed as one exact zero mode (and 15 very light modes corresponding to trace modes) followed by 48 light, constant modes that receive a nonzero eigenvalue due to interactions. They are clearly separated from the rest of the eigenspectrum by a large gap. Furthermore, these low lying modes have a markedly different scaling with volume (see Table~\ref{tab:evfits}). We therefore feel the decision to discard the first 64 eigenvalues from further analysis is well-motivated.

Next we attempt to fit for an exponent $C_n L^{-y_n}$.
We discard the data for $L<5$
from the fit (leaving four data points per coupling choice). This improves the reduced $\chi^2$ value for the
fit, and visual inspection seems to suggest that these data points are
several sigma away, justifying their rejection (see
Figure~\ref{fig:loglogev1}). The exponent $y_n$ tabulated in
Table~\ref{tab:evfits} is therefore a fit to data for $L=5$ to $L=8$ only.

We use a jackknifed least squares fit to the logarithmic data. A jackknife was used for this stage of the
analysis because of its robustness to heteroscedasticity in estimating
fit parameters and their error.

\begin{table}
\begin{center}
\begin{tabular}{|c|c|c|c|c|c|c|}
\hline
 & \multicolumn{2}{|c|}{$\lambda=0.5$} & \multicolumn{2}{|c|}{$\lambda=1.0$} & \multicolumn{2}{|c|}{$\lambda=2.0$} \\
\cline{2-7}
$n$ & $\mu=0.5$ & $\mu=1.0$ & $\mu=0.5$ & $\mu=1.0$ & $\mu=0.5$ & $\mu=1.0$ \\
\hline
9 & $ 3.97 \pm 0.01 $ & $ 4.03 \pm 0.00 $ & $ 3.93 \pm 0.04 $ & $ 4.09 \pm 0.01 $ & $ 3.75 \pm 0.08 $ & $ 3.84 \pm 0.07 $ \\
17 & $ 1.44 \pm 0.04 $ & $ 1.78 \pm 0.05 $ & $ 1.77 \pm 0.28 $ & $ 1.82 \pm 0.09 $ & $ 0.54 \pm 0.38 $ & $ 2.61 \pm 0.12 $ \\
25 & $ 1.52 \pm 0.04 $ & $ 1.80 \pm 0.04 $ & $ 1.83 \pm 0.25 $ & $ 1.90 \pm 0.08 $ & $ 0.95 \pm 0.31 $ & $ 2.71 \pm 0.13 $ \\
33 & $ 2.28 \pm 0.02 $ & $ 2.13 \pm 0.03 $ & $ 2.30 \pm 0.22 $ & $ 2.23 \pm 0.06 $ & $ 1.53 \pm 0.29 $ & $ 3.29 \pm 0.38 $ \\
41 & $ 2.42 \pm 0.06 $ & $ 2.23 \pm 0.03 $ & $ 2.36 \pm 0.26 $ & $ 2.33 \pm 0.07 $ & $ 1.65 \pm 0.30 $ & $ 3.28 \pm 0.42 $ \\
49 & $ 2.16 \pm 0.01 $ & $ 2.09 \pm 0.04 $ & $ 2.22 \pm 0.16 $ & $ 2.22 \pm 0.02 $ & $ 1.36 \pm 0.24 $ & $ 2.93 \pm 0.33 $ \\
57 & $ 2.12 \pm 0.02 $ & $ 2.09 \pm 0.02 $ & $ 2.14 \pm 0.12 $ & $ 2.15 \pm 0.01 $ & $ 1.13 \pm 0.24 $ & $ 2.75 \pm 0.28 $ \\
65 & $ 1.76 \pm 0.03 $ & $ 1.75 \pm 0.01 $ & $ 1.83 \pm 0.04 $ & $ 1.73 \pm 0.02 $ & $ 1.07 \pm 0.23 $ & $ 2.61 \pm 0.26 $ \\
73 & $ 1.77 \pm 0.03 $ & $ 1.75 \pm 0.02 $ & $ 1.85 \pm 0.03 $ & $ 1.73 \pm 0.03 $ & $ 0.99 \pm 0.17 $ & $ 2.37 \pm 0.23 $ \\
\hline
\end{tabular}
\end{center}
\caption{\label{tab:evfits} Results of fitting $\langle\lambda_n(L)\rangle$ to $C_n L^{-y_n}$ for
several $n$ at various $\lambda$ and $\mu$ ($n=1$ is an exact zero mode); the values of $y_n$ are tabulated. The fits exclude data for
$L<5$. Note that the results for $\lambda_{65}$ can be
compared directly with earlier plots.}
\end{table}

\subsection{Attempting to measure the anomalous mass dimension}\vspace{-0.1in}
As we are limited to rather small volume it may seem optimistic to
hope that one can obtain an estimate for the anomalous mass dimension
from these measurements, but a very crude estimation is still possible
following the method of Ref.~\cite{Patella:2012da}. Better results
along the same lines would be easy to obtain if data for larger
volumes were readily available.

The basic quantity is the integrated eigenvalue density that yields
the mode number per unit volume
\begin{equation}
{\bar \nu}(\Omega) = \int_0^\Omega d\lambda \; \rho(\lambda); \qquad
\rho(\lambda) = \lim_{V\to\infty} \sum_k \left< \delta(\lambda-\lambda_k)\right>.
\end{equation}
In the thermodynamic limit the sum over Dirac delta functions could be
interpreted literally, but at finite volume the measure is necessarily
less sharp.

\begin{figure}
\begin{center}\includegraphics[trim=1 1 1 1,clip=true,width=0.6\textwidth]{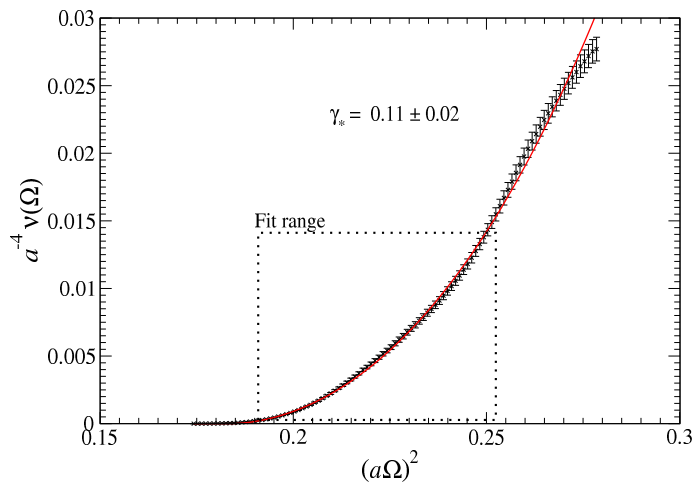}\end{center}
\caption{\label{fig:patella} Plot of the integrated eigenvalue density
  with $\lambda=\mu=0.5$ for $L=8$. The fitted region is shown. A bootstrap error analysis yields an estimate of $\gamma_* = 0.11 \pm 0.02$ for this case, with $\chi^2/\mathrm{d.o.f.} \approx 0.99$.}
\end{figure}

We show the result of calculating ${\bar \nu}(\Omega)$ in
Figure~\ref{fig:patella} for $L=8$ at $\lambda=0.5$,
$\mu=0.5$. The finite volume effects on the eigenvalue spectrum are
already quite clear in this plot for $\omega>0.27$, justifying our decision
only to calculate the lowest-lying 200 eigenvalues; larger lattices
would need to use more sophisticated methods such as the projection
technique discussed in Ref.~\cite{Patella:2012da}.

Despite having limited data, we continue. We repeat the
analysis of Ref.~\cite{Patella:2012da} by attempting to fit the ansatz
\begin{equation}
{\bar \nu}(\Omega) = {\bar \nu}_0 + A\left[ \Omega^2 - m^2
  \right]^{\frac{2}{1+\gamma_*}}
\end{equation}
to our data for the integrated eigenvalue density. We take ${\bar
  \nu}_0 = 0$ as fitting with this as a parameter leads to results
consistent with zero. In any case, Figure~\ref{fig:patella} suggests
that there are no low eigenmodes that we can exclude in this manner.

A nonlinear least squares fit is carried out to determine $A$, $m$ and $\gamma_*$,
using the data shown in
Figure~\ref{fig:patella}. Note that the error bars for the integrated
eigenvalue density at each point are, of course, correlated. With that
in mind we first systematically scan all ranges of $\Omega$ for that
yielding a reduced chisquare closest to unity, then vary the lower and
upper ranges of the fit separately to try and locate a `plateau'
nearby.

The results of this procedure yield a value
$\gamma_*=0.11\pm0.02$. A consistent result is obtained carrying out the same analysis on our $L=7$ configurations. Therefore, despite the small volume, it is possible to use the integrated eigenvalue density in this context to measure -- at least crudely -- the mass anomalous dimension. Given the limited volume we would argue that this result is consistent with an expectation that $\gamma_*$ would be zero in a more comprehensive study.

\section{Conclusion}\vspace{-0.2in}
We have carried out a study of the behaviour of the Dirac eigenspectrum for a lattice implementation of $\mathcal{N}=4$ SYM. The low-lying eigenvalues have a very different structure that we interpret as being due to the approximate zero modes and trace modes that remain in the theory.

A preliminary calculation of the integrated spectral density was carried out, and we see that the mass anomalous dimension is very small and consistent with zero. Future work will require studies at substantially larger volumes to put this analysis on a more robust footing. These calculations are underway.

\section*{Acknowledgement}\vspace{-0.2in}
This work was supported by the U.S. Department of Energy grant under
contract no. DE-FG02-13ER41985 and a DARPA Young Faculty
Award. Simulations were performed using USQCD resources at Fermilab,
at the Niels Bohr Institute and on the Finnish Grid
Infrastructure. DJW would like to thank Syracuse University for its
hospitality and acknowledges support from Academy of Finland grant
1134018.

\end{document}